\documentclass[aps,twocolumn,showpacs,pre,preprintnumbers,amsmath,amssymb]{revtex4-1}
\pdfoutput=1
\usepackage{graphicx}
\usepackage{dcolumn}
\usepackage{bm}

\begin{document}
\title{Aggregation of frictional particles due to capillary attraction}%

\author{Marie-Julie Dalbe}
\author{Darija Cosic}
\author{Michael Berhanu}
 \altaffiliation[Present address : ]{Mati\`ere et Syst\`emes Complexes (MSC), Universit\'e Paris Diderot, CNRS (UMR 7057), 75013 Paris, France}
\author{Arshad Kudrolli}

\affiliation{%
Department of Physics, Clark University, Worcester, Massachusetts 01610, USA
}%

\date{\today}

\begin{abstract}
Capillary attraction between identical millimeter sized spheres floating at a liquid-air interface and the resulting aggregation is investigated at low Reynolds number. We show that the measured capillary forces between two spheres as a function of distance can be described by expressions  obtained using the Nicolson approximation at low Bond numbers for far greater particle sizes than previously assumed. We find that viscous hydrodynamics interactions between the spheres needs to be included to describe the dynamics close to contact. We then consider the aggregates formed when a third sphere is added after the initial two spheres are already in contact. In this case, we find that linear superposition of capillary forces describes the observed approach qualitatively but not quantitatively. Further, we observe an angular dependence of the structure due to a rapid decrease of capillary force with distance of separation which has a tendency to align the particles before contact. When the three particles come in contact, they may preserve their shape or rearrange to form an equilateral triangle cluster - the lowest energy state - depending on the competition between attraction between particles and friction. Using these observations, we demonstrate that a linear particle chain can be built from frictional particles with capillary attraction. 
\end{abstract}


\maketitle

\section{Introduction} 
Aggregates can be observed to form in particulate systems with attractive interactions with shapes that depend on the nature of forces between particles and on their initial positions. An important example is the aggregation of floating objects at the surface of a liquid due to capillarity. This phenomenon can be easily observed in everyday examples such as clustering of bubbles in a sink~\cite{Nicolson}, clumping of  breakfast cereals floating in a bowl filled with milk~\cite{Vella}, and biomaterials such as pollen or eggs of some insects species observed floating at the surface of ponds~\cite{Singh3}, and even swimming nematodes~\cite{gart11}. Capillary aggregation has many important applications as in flotation processes in ore extraction, and self assembly of micron-sized floating particles to fabricate new 2D-materials~\cite{Bowden}. Further, this phenomenon has been exploited to study formation of ramified fractal aggregates~\cite{Allain1986}, and recently by our group to examine the heterogeneous nature of cohesive granular media using spheres floating at a liquid-air interface~\cite{Berhanu}. 
While floating particles of the same kind always attract each other, particles with different wetting properties can repeal each other. For simplicity, we limit our discussion in the following to identical spherical particles.

The first attempt to introduce capillary forces due to a liquid interface between floating bodies, was given by Poynting and Thomson~\cite{Poynting}, who derived the forces between semi-immersed plates. The level of fluid in between is increased if the plates are hydrophilic or decreased if the plates are hydrophobic. In both cases the curvature of the interface modifies hydrostatic pressure between the plates, which overcomes the pressure on the external sides, leading to an attraction between the plates. This mechanism cannot be directly applied to floating particles~\cite{Vella}, because gravitation, buoyancy and capillary forces set the height of the objects. Because the location of the contact line and the shape of the meniscus results from a competition between gravity and capillarity, its horizontal extension is of order of the capillary length $L_c=\sqrt{\gamma / (\rho \, g)}$, where $\gamma$ is the surface tension between the liquid and the atmosphere, $\rho$ the density of the liquid, and $g$ the gravitational acceleration. Thus, a particle in range of the meniscus caused by other particles, is out of equilibrium because the horizontal projection of surface tension and hydrostatic pressure is no longer isotropic. However, the exact calculation of capillary interactions becomes difficult even for spherical particles due to geometrical complexity. Therefore, an analytical approach becomes possible only by simplifying the problem. 

Nicolson~\cite{Nicolson} first proposed a linear superposition approximation to calculate the force between two identical bubbles at a liquid interface, which has since been applied to floating spheres~\cite{Chan,Vella}. 
Using this approximation implies that the calculations are restricted to small deformations of the interface and to particle of radius $R$ which is small compared to $L_c$. The Bond number $B_o=R^2 / {L_c}^2$ is the corresponding dimensionless parameter to compare particle size and the capillary length and consequently linear calculations are limited to $B_o \ll 1$. The force also depends on the contact angle $\theta$ at the contact line between the atmosphere, the fluid and the particles. A complementary approach starting with Young-Laplace equation in bipolar coordinates for small deformations and particle sizes has been also used to derive the shape of the liquid interface and the force of attraction~\cite{Kra1,Kra2,Paunov}. 
 Singh {\em et al.}~\cite{Singh,Singh3} have developed a numeric simulation to study the motion of floating bodies coupled with free surface flow, but the results were not compared with experimental measurements. 
Capillary forces between partially submerged spheres has been experimentally investigated but the particles were constrained from moving~\cite{Velev,Dushkin}, or with sub-millimeter sized particles at liquid-liquid interface where Nicolson approximation is expected to work~\cite{Vassileva}. 

For  particles larger than a millimeter, the approximations used in theoretical derivations discussed above become less obvious. The capillary force between millimeter sized spheres has been measured by Camoin {\em et al.}~\cite{Camoin} and a decreasing exponential shape was found. But the particles in that study were not free to move and no comparison with theory was presented. 
A further important factor which has received little attention during the aggregation is the surface friction of the particles which can become important when more than two particles aggregate~\cite{Berhanu}. Therefore, careful experiments are necessary to clarify the physics of capillary aggregation for millimeter sized particles. 

In this paper, the force of attraction between floating spheres is investigated by measuring and analyzing trajectories of identical spheres with friction. In particular, we investigate if the mechanism of aggregation that have been calculated and tested for $R << L_c$ can be extended to $R \sim L_c$. We will also examine dynamics of three particles to study the effect of friction and validity of linear superposition of capillary forces for many particle systems. 
We introduce first the theoretical background related to the capillary interactions and the effect of viscous drag in Sec.~\ref{sec:theory}, and then describe the experimental apparatus in Sec.~\ref{sec:apparatus}. The experimental study of the dynamic of two initially isolated particles and the corresponding analysis are presented in Sec.~\ref{subsec:2parts}, experiments with three particles are discussed in Sec.~\ref{subsec:3parts}. We conclude by demonstrating fabrication of a linear chain of spheres by exploiting friction, and some remarks on the general implication of our study.

\section{Background}
\label{sec:theory}
To discuss the nature of approximations, we first describe in brief the derivation of the force of attraction between two identical floating spheres using the Nicolson approximation~\cite{Nicolson,Chan,Vella}. Then, we consider the viscous interactions that need to be calculated to describe particle dynamics near contact. 

\subsection{Capillary attraction between floating spheres}
\begin{figure}[t]
\centering
\includegraphics[width=8.5cm]{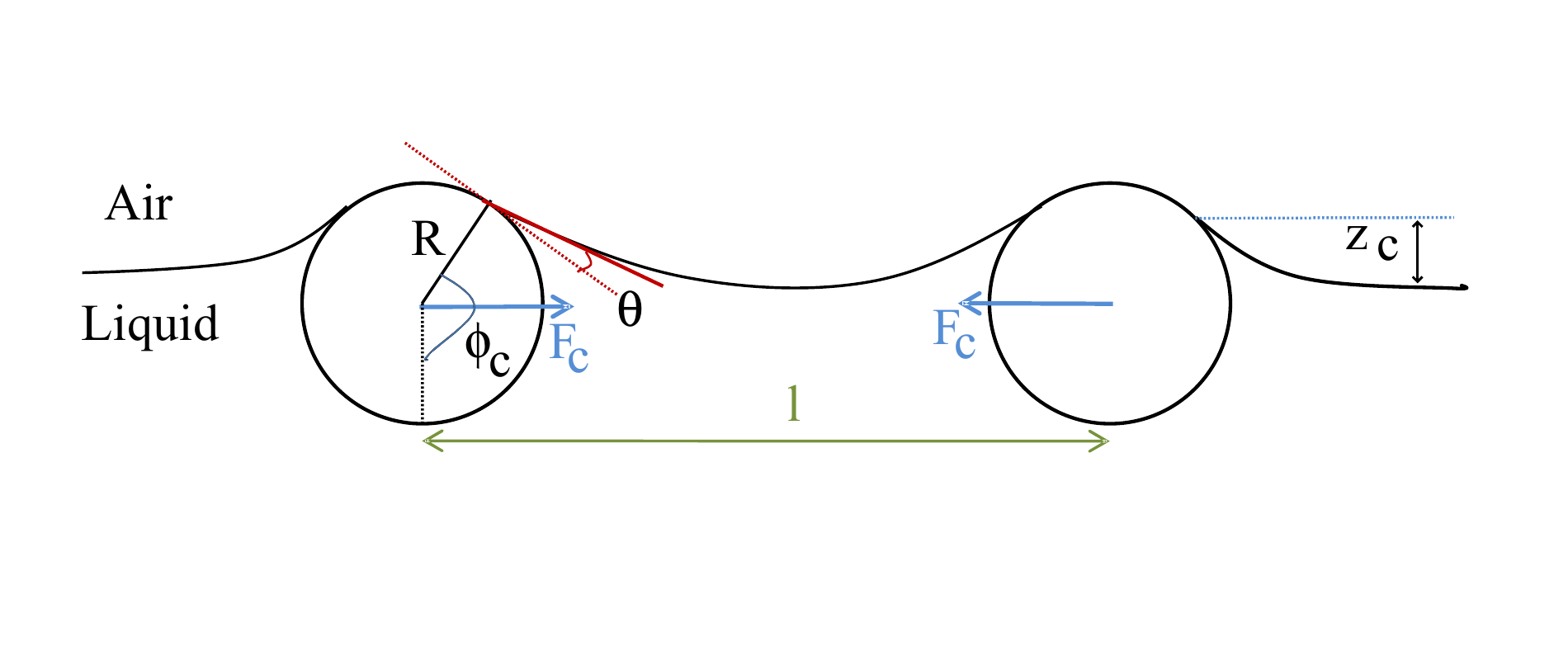}
\caption{A schematic diagram of two spherical particles floating at a liquid-air interface.}
\label{fig:geom2part}
  \end{figure}

A schematic of spheres with density $\rho_s$ floating at a liquid-air interface and its deformation is shown in Fig.~\ref{fig:geom2part}. We denote the height of the contact line relatively to the fluid level by $z_c$, and the interfacial slope at the contact line by $z_{c}'$. The vertical position of the sphere is parameterized by the angle $\phi_{c}$ between the lowest point of the sphere and the level of the contact line. From geometrical considerations this angle is related to the contact angle by the relation: 
$$\phi_{c} =  \pi - \theta + \mathrm{arctan} (z_{c}') $$
To find $\phi_{c}$, we use the vertical force balance for a particle at rest at a liquid-air interface 

\begin{equation}
 P+A+ T_z = 0 \,\, .  
\label{eq:balance}
\end{equation}
Where,   ${P}$ is the sphere weight, ${A}$ the buoyancy for a semi-immersed body, and $T_z$
the surface tension integrated along the contact line which by symmetry is along the vertical direction and are given by
\begin{eqnarray}
P & = &- 4/3 \, \pi \, R^3 \, \rho_s \, g \,\, ,  \nonumber\\
A & = & \pi \,R^3 \, \rho \, g  \left( \dfrac{2}{3} \left(1-{\cos}^3 \phi_c \right) - \left(\dfrac{z_c}{R} + \cos \phi_c \right) {\sin}^2 \phi_c   \right)\,\, ,\nonumber\\ 
{\rm and} \nonumber\\
T_z &=& 2 \pi \, \gamma \, R \sin \phi_c  \dfrac{z'_c}{(1+{z'_c}^2)^{1/2}} \,\, . \nonumber
\end{eqnarray}
Assuming that  $z_{c}' \ll 1$ and $z_c \ll R$, the following expression can be obtained~\cite{Chan,Vella}:
\begin{equation}
\mathrm{sin}\phi_{c} \, z'_c = \mathrm{sin}\phi_{c} \, \mathrm{tan}(\phi_{c}+\theta)=B_o \, \Sigma  \,\, .
\label{eq:slope}
\end{equation}
Here, $\Sigma= \dfrac{2 \rho_{s}/\rho -1}{3}- \dfrac{1}{2} \cos \theta +\dfrac{1}{6} \cos^{3} \theta $.
Given the physical parameters of the particle and fluid, the value of $\phi_{c}$ and  $z_{c}'  = \tan ( \phi_c + \theta) $ can be then numerically computed.
Because of axisymmetry, $z$ as a function of the distance to the particle center $r$ is given by the Young-Laplace relation,  
$\dfrac{z(r)}{{L_c}^2}=\mathcal{C} (r)$, where $\mathcal{C}(r)$ is the local interface curvature. For small $z'(r)$, we then obtain 
\begin{equation}
\dfrac{z(r)}{{L_c}^2}= \nabla^2 z(r) \,\, .
\label{eq:eq}
\end{equation}
This equation with boundaries conditions $z \rightarrow 0 $ as $r\rightarrow \infty$ and $z'(r=R \mathrm{sin} (\phi_c))=z'_c$ admits solution with modified Bessel function of the first kind $K_n (x)$, and further noting that $\dfrac{d K_0 (x) }{d x} = - K_1 (x)$, one obtains
\begin{equation}
z(r)=- \tan (\phi_{c}+\theta)\, L_{c} \dfrac{K_{0}(r/L_{c})}{K_{1}(R \, \sin (\phi_{c})/L_{c})}\,\, . 
\label{eq:surface}
\end{equation}
The numerical value of the deformation at a point along the contact line (meniscus size) $z_c$ is found by replacing the value of $r$ by $R \, \sin \phi_c $. 

Next, we use the Nicolson approximation~\cite{Nicolson,Chan} to compute the force experienced by a particle due to the meniscus of a second particle  (see Fig.~\ref{fig:geom2part}). In this case, the horizontal projection of the capillary force integrated along the contact line has a resultant which is directed towards the center of the second sphere. The expression remains difficult to solve. However, using linear superposition,  and noticing that for small deformations, the resultant of surface tension is modified in direction but not in amplitude, $F_{cap}$ can be approximated as:
$F_{cap}=T \, {z'_2}$, where $z'_2=\dfrac{d z_2}{d l}$ is the slope of interface created by the second particle~\cite{Allain1992}. Using Eq.~(\ref{eq:balance}) and expressing the elementary work of the capillary force, we obtain: $$ \delta W = - F_{cap}\, {d l} = (P+A) \, d z_2. $$ By integration and using Eq.~(\ref{eq:slope}), the energy of interaction, $E(l)$ between these particles is expressed as the product of an effective weight $P_{eff}=(P+A)=2\pi \, \gamma \, R\, B_o \, \Sigma $ and the liquid surface deformation $z(r=l)$.
Using the expression of $\Sigma$ given by Eq.~(\ref{eq:slope}), the effective weight can be expressed as:  $$ 
 P_{eff}=2\pi \, \gamma \, R\,\sin ( \Phi_{c} ) \, \mathrm{tan}(\Phi_{c}+\theta)
 $$
  Then the energy of interaction $E(l) \,= \,  P_{eff} \, z(r=l)$ is written as:
$$
E(l)  =  - 2\pi \, \gamma \, R\, \dfrac{\sin ( \Phi_{c}) \,  {\tan (\Phi_{c}+\theta)}^2\,  L_{c} \,K_{0}(l/L_{c})}{K_{1}(R \, \sin (\Phi_{c})/L_{c})} 
$$

Substituting in $F_{c}(l)=-dE/dl$, and using Eq.~(\ref{eq:surface}), we obtain the following expression for the capillary force between two particles:
\begin{eqnarray}
F_{c}(l) & = & - C_{VM} \, K_{1}(l/L_{c}) \,\, ,
\label{eq:forceVella}
\end{eqnarray}
where,
$ C_{VM}  =  2 \pi\, \gamma\, R   \sin(\phi_{c})\dfrac{(\tan(\phi_{c}+\theta))^2}{K_{1}(R\, \sin(\phi_{c})/L_{c})} \,\,.$
For $l \gg L_{c}$, we can use the asymptotic form of $K_1$ : $K_1(x) \approx \sqrt{\dfrac{\pi}{2x}}e^{-x}$ for $x \gg 1$. 
Thus $F_{c}$ decreases rapidly for distances larger than the capillary length. $F_{c}$ is also attractive and this feature can be explained following Singh and Joseph~\cite{Singh}. For light hydrophilic particles, rise of liquid between particles decreases the slope of liquid interface at the contact line. The horizontal projection of the tension force is therefore increased between particles, exceeding those on external side and leading to an attractive force. The argument remains the same for heavier hydrophobic particles, by decreasing the slope of liquid interface at the contact line.

It may be noted that a similar expression valid for small $B_o$ is found using a different method~\cite{Vassileva,Paunov,Kra1,Kra2}, but with a slightly different prefactor:
\begin{eqnarray}
F_{c'}(l) & = & - C_{Pa} \, K_{1}(l/L_{c}) \,\, ,
\label{eq:forceVassileva}
\end{eqnarray}
with, $C_{Pa} = 2\,\pi \,\gamma \, L_{c} \dfrac{(\tan(\phi_{c}+\theta))^2}{(K_{1}(R\,\sin(\phi_{c})/L_{c}))^2}\,\,.$

For $R\ll L_{c}$, the prefactors can be simplified, and one can replace the Bessel function by its asymptotic from $K_{1}(x)\approx 1/x$ for $x \ll 1$. Both Eqs. (\ref{eq:forceVella}) and (\ref{eq:forceVassileva}) then simplify to: 
\begin{eqnarray}
F_{c0}& = & - C_0 \, K_{1}(l/L_{c})\,\, ,
\label{eq:force}
\end{eqnarray}
with $C_0 = 2\, \pi\, \gamma\, R\, {B_o}^{5/2} \, \Sigma ^{2}$. It is important to note that in this equation, the factor before the Bessel function was mainly determined by the hypothesis $B_o\ll 1$, whereas the condition of validity of Eq.~(\ref{eq:forceVella}) is given mainly by $z_{c}' \ll 1$. Moreover in using the Nicholson approximation, contribution of hydrostatic pressure is completely neglected which may not be true for millimeter sized spheres.  While the work of Allain and Cloitre for cylinders~\cite{Allain1992} shows that pressure contribution becomes negligible when $B_o < 10$, an estimate of its amplitude has not been reported for millimetric spheres when $B_o$ is not small. 
Finally, it can be noted that the expression for the capillary force between two spheres expressed in Eqs. (\ref{eq:forceVella}), (\ref{eq:forceVassileva}) and (\ref{eq:force}) is a product of a Bessel function $ - K_{1}(l/L_{c})$ giving the spatial dependency and a constant depending on the model ($C_{VM}$, $C_{VA}$ and $C_0$). In the following the constant is simply labelled $C$, regardless of the theoretical model.  

These results can be also applied to the case with more than two particles. Because equation Eq.~(\ref{eq:eq}) is linear, deformation felt by one particle is the sum of those created by the other particles individually. The resulting capillary potential energy is again obtained by the Nicolson approximation, \textit{i.e.} by multiplying the effective weight by the fluid surface deformation. Consequently the capillary force felt by one particle is also the sum of the force created by the other particles and computed using Eq.~\ref{eq:forceVella}. 

\subsection{Hydrodynamics of floating spheres}
\label{subsec:balance}
Because a floating sphere begins to move due to capillary interaction,  one has to consider additional hydrodynamic interactions. We limit our analysis to the case of small Reynolds number $R_e=\frac{\rho L V}{\mu}$ to neglect inertial effects and also small Capillary numbers $C_a=\frac{\mu V}{\rho \gamma}$ to neglect motion of the contact line on the particle, where $L$ is a typical length scale, $V$ a typical velocity, and $\mu$ the dynamic viscosity. The drag force for a partially immersed sphere is given by the Stokes law corrected by a drag coefficient $f_{d}$~: 
\begin{equation}
\mathbf{F_{d}}= - 6  \pi \, \mu \, R \, f_{d} \, \mathbf{v},
\label{eq:drag}
\end{equation}
where,  $\mathbf{v}$ is the particle velocity. $f_{d}$ depends on the vertical position of the sphere relative to the interface, the contact angle, surface tension and density of particle and liquid. 
In order to take into account the hydrodynamic interactions due to the flow created by the second sphere, we adopt the concept of hydrodynamic mobility introduced by Batchelor~\cite{Batchelor1976} for colloidal particle motion. For low $Re$, the difference of the capillary and drag forces projected along the axes between particles centers leads to: 
\begin{eqnarray}
0 &=& F_{c,1}+F_{d,1}-F_{c,2}-F_{d,2} \,\, , \nonumber \\
0 & = & -2C\, K_1(l/L_c)-6 \, \pi \, \mu \, R \, f_{d} \, \dfrac{dl}{dt} \,\, ,
\end{eqnarray}
or 
\begin{equation}
-\dfrac{dl}{dt}=\dfrac{C\, K_1(l/L_c)}{3 \, \pi \, \mu \, R \, f_{d}}.
\label{eq:dyn2}
\end{equation}
This last expression is corrected by multiplying it by the hydrodynamic mobility $G(x)$ in terms of $x = l/R$ for two spheres along the line join their centers~\cite{Batchelor1976}, and is given by~\cite{Batchelor1976,Batchelor1972,Vassileva}:
\begin{equation}
G(x)=1-\dfrac{3}{2 x}+\dfrac{1}{x^3}-\dfrac{15}{4 x^4}-\dfrac{4.46}{1000}(x-1.7)^{(-2.867)}\,\, .
\label{eq:G}
\end{equation}
Therefore, the equation of the motion of a sphere is given by
\begin{eqnarray}
\dfrac{dl}{dt} &= & K_{G}\, G(x)\, K_{1}(xR/L_{c})\,\, ,
\label{eq:balanceG}
\end{eqnarray}
where, $K_{G} =  \dfrac{2 \, C}{6  \pi \, \mu \, f_{d}\, R}.$ 
This expression predicts no hydrodynamic interactions when the spheres are far from each other ($G(x\rightarrow \infty)=1$) and $\dfrac{dl}{dt} = 0$ when the spheres come in contact ($G(2)=0$), because of lubrication created by the liquid between the spheres.   

These results show that important features of dynamics of capillary aggregation can be extracted from the analysis of particles trajectories.
In section~\ref{subsec:2parts}, we will study examples with two floating spheres with $B_o \sim 1$ and compare experimental results with Eq.~(\ref{eq:balanceG}).

  \section{Experimental setup}
  \label{sec:apparatus}
\begin{figure}[t]
\begin{center}
\includegraphics[width=8.5cm]{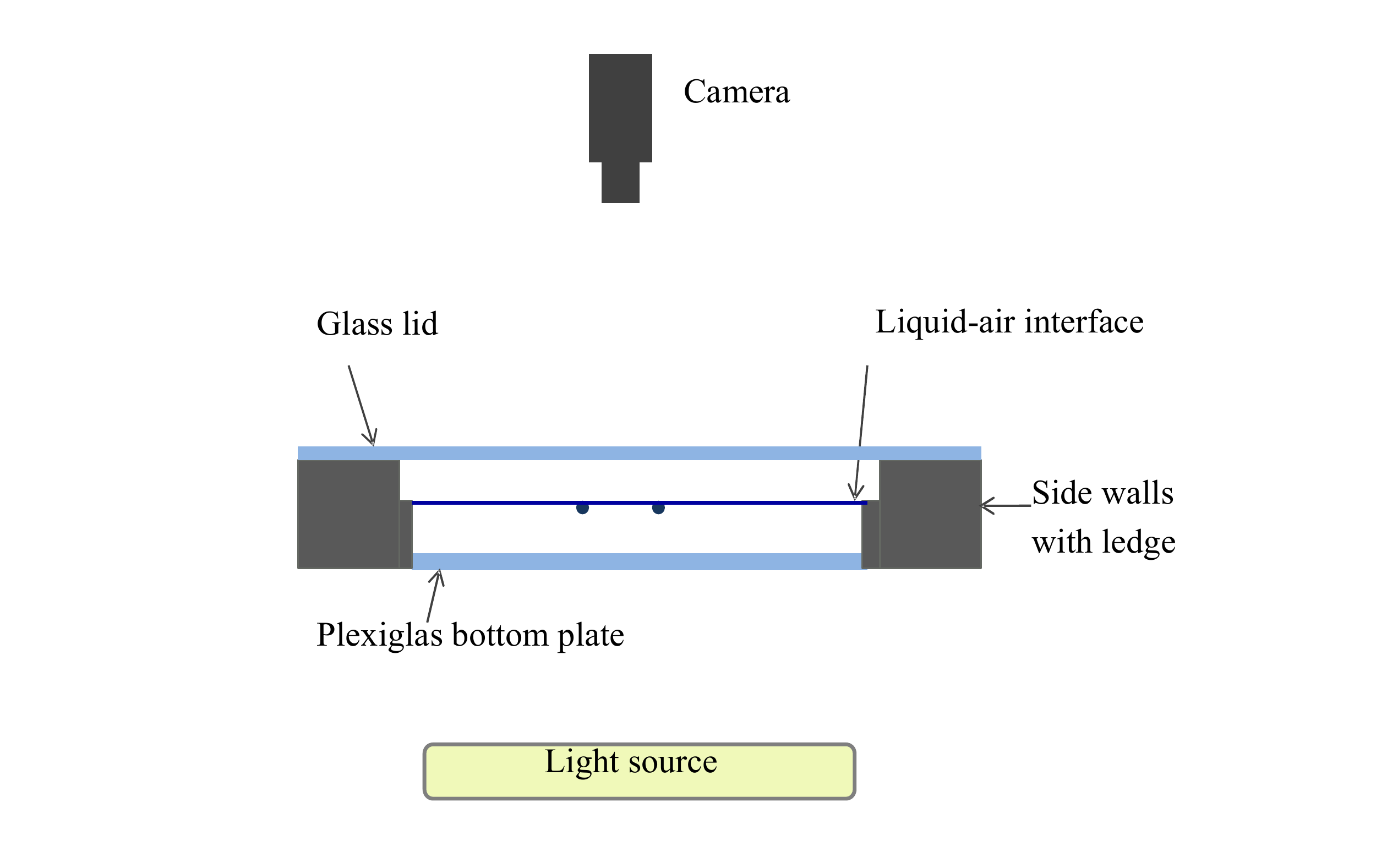}
\end{center}
\caption{A crossection of the experimental apparatus with two particles floating at the liquid-air interface. The liquid is filled up to a ledge in the sidewall to minimize particle interactions with the container.}
\label{fig:setup}
\end{figure}

Experiments are conducted in a container with an aluminum frame with dimensions measuring $20.3$\,cm long, $20.3$\,cm wide, and $2.54\,$cm deep, and a Plexiglas bottom (Fig.~\ref{fig:setup}). A glass lid is placed above to avoid evaporation and prevent dust from falling in. To avoid boundary effects, meniscus is pinned at a ledge along the sidewalls of the container in order to minimize the meniscus and ideally obtain a flat surface. The container is kept horizontal to within $0.1$\,degrees, and the level of liquid can be adjusted with a syringe. 
The particles are observed from above with a CMOS $1280 \times 1024$ pixels camera, with a telecentric lens, and back lit through the transparent bottom boundary. Images are recorded with a frame rate of one image per second, and particles are tracked with standard algorithms implemented in \textsc{ImageJ}. Relative error in finding particles center is less than 2$\%$ of the diameter.    
  
\begin{figure}
\centering
\includegraphics[width=5cm]{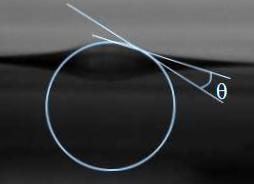}
\caption{A side view of a floating particle with $D = 3.175$\,mm. The contact angle $\theta$ corresponds to the angle between the liquid-air interface and the tangent to the particle at the same point. }
\label{fig:wettingangle}
\end{figure}

In order to obtain a sufficiently viscous and dense liquid, we use a mixture of glycerol ($90\%$ by weight) and distilled water. The physical properties of this liquid are summarized in Table~\ref{tab:prop} using the data from Ref.~\cite{Phys}.
Polyethylene (High-density polyethylene) or nylon (Nylon 6) spheres are used in our study summarized in the Table~\ref{tab:propsphere}. These materials are hydrophilic and lighter than the liquid, and were chosen in order to form stable aggregates. 
The contact angle $\theta$ needed to compare with calculations is estimated directly by imaging a sphere floating on the liquid from the side. The particles are first immersed deep into the liquid and then allowed to reach their vertical equilibrium. 
The contact angle is measured from the image, as shown on Fig.~\ref{fig:wettingangle} and values are reported in Table~\ref{tab:propsphere}. Using this data, we estimate the slope of the liquid interface $z_c'$  and the amplitude of meniscus
$z_c$, using formulae in Eqs.~\ref{eq:eq} and \ref{eq:slope}. Even for millimetric sized particles where Bond number is of order one, approximations used in Section~\ref{sec:theory}, \textit{i.e.} $z_c/R \ll 1$ and $z_c' \ll 1$, are satisfied. Furthermore, we notice that the vertical position of the sphere center $z_{center}=z_c + R\,\mathrm{cos} \phi_c$ is negative, so particles centers are below the surface, which allows us to assume that the drag coefficient for semi-immersed sphere $f_d \approx 1$. Also when particles are touching each other, the contact point between particles is located below the liquid surface, which prevents high interface distortion when particles are close.

Finally, we note that experiments are very sensitive to vibrations and thermal convection.  Therefore, room temperature is regulated at $23^\circ$\,C and several hours are needed after filling the container to reach thermal equilibrium to commence experiments.

\begin{table*}
\centering
\begin{tabular}{cccc}
\hline
Density $\rho$ (kg\,m$^{-3}$) & Surface tension $\gamma$ (N\,m$^{-1}$) & Viscosity $\mu$ (Pa\,s) & Capillary length $L_{c}$ (mm) \\ \hline \hline

1233 $\pm$ 3& 0.064 $\pm$ 0.002 & 0.175 $\pm$0.015 &  2.30 $\pm$ 0.04 \\ \hline 
\end{tabular}
\caption{Physical properties of glycerol-water mixture used in the experiments~\cite{Phys}.}
\label{tab:prop}
  \end{table*}
  
\begin{table*}
\centering
\begin{tabular}{ccccccccc}
\hline
Material & $D$ (mm) & $\rho_s$ (kg m$^{-3}$) & $\theta$ (degrees) & $B_o$ & $\phi_c$ (degrees) & $z_c$ (mm) & $z_c'$ & $z_{center}$ (mm) \\ \hline
\hline
Polyethylene & 6.35 $\pm$ 0.05 & 950 &  40 $\pm$ 3 & 1.91 $\pm$ 0.1 & 124 & 0.49 & - 0.29 & -1.27 \\  
Polyethylene & 3.175 $\pm$ 0.05 & 950 &  20 $\pm $ 2 & 0.48 $\pm$ 0.03 & 151 & 0.16 & - 0.15 & -1.23 \\  
Nylon & 3.175 $\pm$ 0.03 & 1150 & 5 $\pm$ 5  & 0.48 $\pm$ 0.03 & 169 & 0.074 & - 0.11 & -1.48 \\  
Nylon & 2.38125 $\pm$ 0.05 & 1150 & 5 $\pm$ 5  & 0.27 $\pm$ 0.02 & 171 & 0.038 & - 0.075 & -1.14\\  
Nylon & 1.5875 $\pm$ 0.05 & 1150 & 5 $\pm$ 5 & 0.12 $\pm$ 0.009 & 173 & 0.014 & - 0.042 & -0.77 \\ \hline 
\end{tabular}
\caption{The diameter $D$ and the density $\rho_s$ of the particles were given by the provider, the contact angle $\theta$ is measured directly (see Fig.~\ref{fig:wettingangle}), and the Bond number $B_o$ is the ratio ${R^2}/{L_c}^2$. The other parameters characterize the equilibrium position of a floating sphere including the angle $\phi_c$ computed from Eq.~(\ref{eq:slope}). Amplitude of the meniscus around the particles $z_c$ and meniscus slope $z_c'$ are estimated using Eqs. (\ref{eq:eq}) and (\ref{eq:slope}). Vertical position of sphere center is given by $z_{center} = z_c + R\,\mathrm{cos} \phi_c$. }
\label{tab:propsphere}
  \end{table*}

\section{Attraction between two floating particles}
\label{subsec:2parts}
\begin{figure}
\centering
\includegraphics[width=8.5cm]{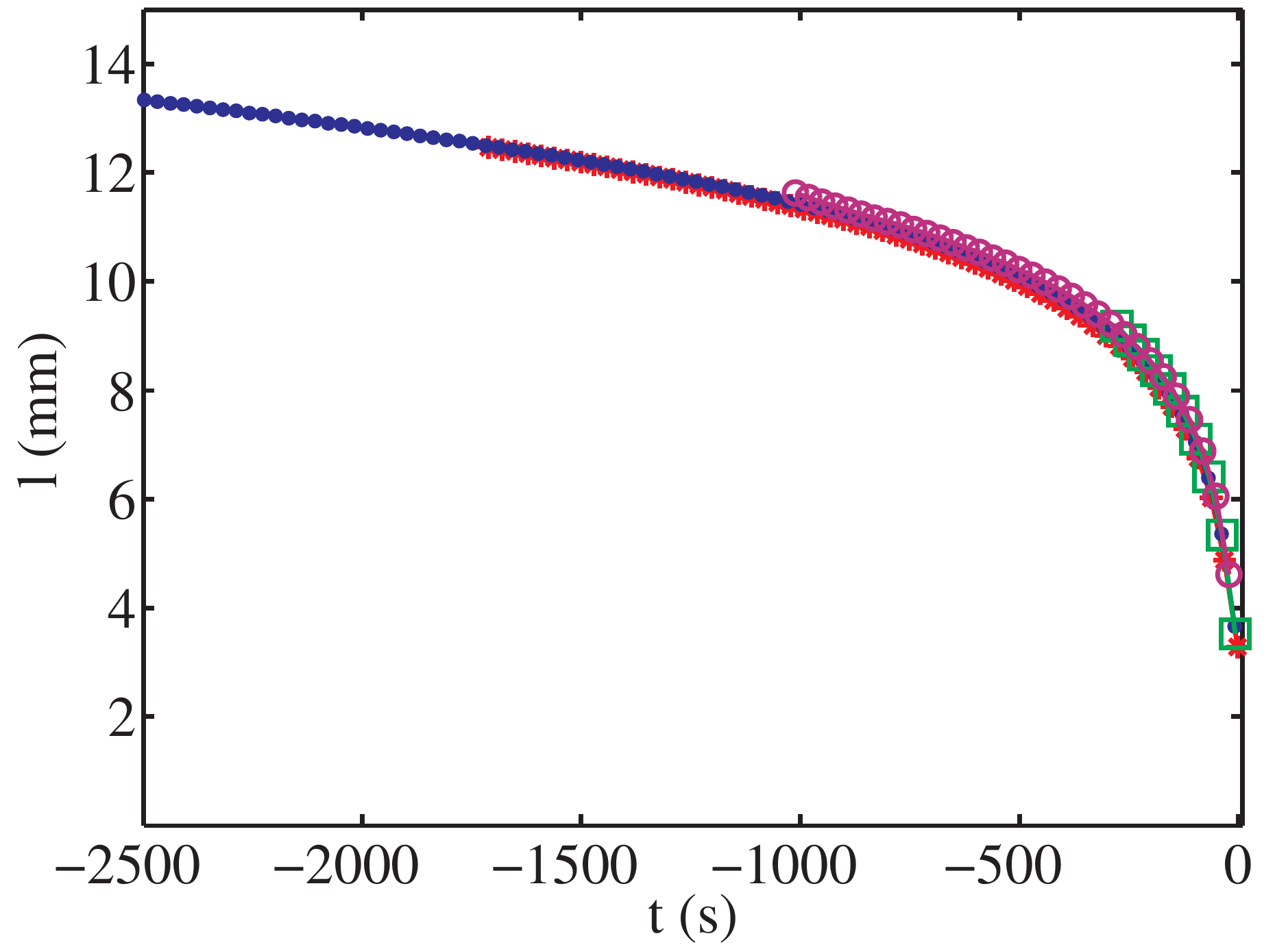}
\caption{The distance between the centers of the two polyethylene particles versus time ($D=3.175$\,mm). The time is chosen to be equal to zero when the spheres come into contact. The different symbols correspond to various initial separation distance. 
}
\label{fig:distancevstime3mm}
\end{figure}

In this section, we discuss the motion of two initially separated floating particles. 
To obtain reproducible data, the particles are first totally immersed in the liquid and then placed in their initial position with negligible velocities. 
We plot the distance between the centers of two polyethylene spheres $l$ as a function of time in Fig.~\ref{fig:distancevstime3mm}. Curves for various initial positions overlap with each other when we chose the time of origin to be the time of contact. This curve shows that the particle dynamics before contact, is a function only of separation distance. 

\begin{figure}
\centering
\includegraphics[width=8.5cm]{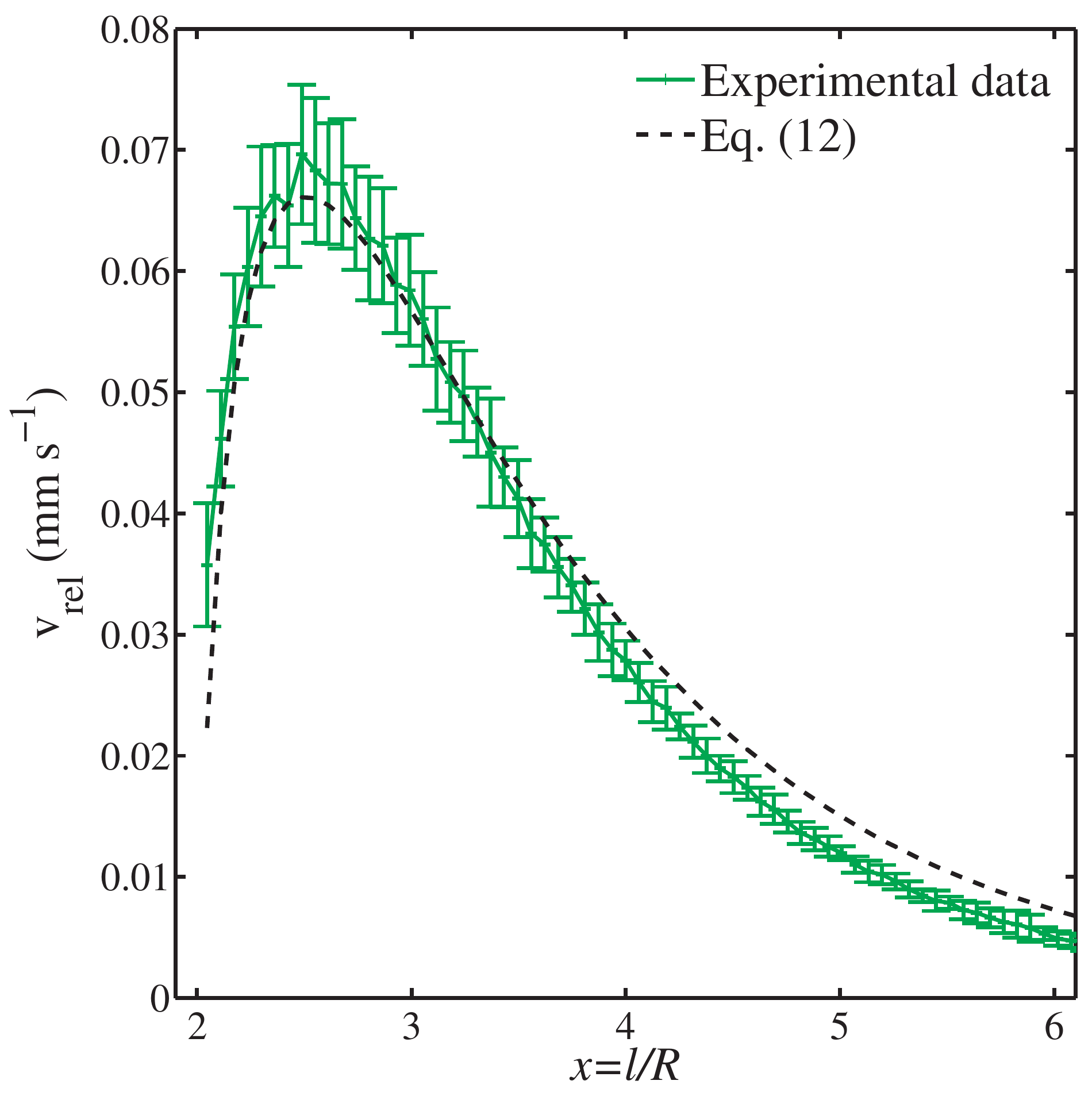}
\caption{Approach velocity $v_{rel}$ versus $x=l/R$ of Polyethylene spheres with $D=3.175$\,mm. Errors bars are estimated from the statistical dispersion using 12 data sets. The curve is fitted by equation Eq.~(\ref{eq:balanceG}) with  coefficient $K_{G}=9.10 \times 10^{-4} \pm 0.1 \times 10^{-4} m.s^{-1}$.}
\label{fig:velocityfit1}
\end{figure}

We then determine the velocity of the spheres $v_{rel}=dl/dt$ from the separation distance over time and plot it in Fig.~\ref{fig:velocityfit1} versus $x=l/R$. The data was averaged over ten experiments to reduce statistical and measurement errors. The velocity is observed to first increase rapidly as the spheres approach each other, and then decrease as the particles approach each other because of lubrication, and reaches zero once particles come in contact due to hard core repulsion. 

From this measured velocity, we can extract the drag force experienced by the particles, and assuming that inertia is negligible, we can also obtain the capillary force by equating it with the drag force. 
Now the maximum velocity is approximately $v_{rel}/2 \sim 4 \times 10^{-5}$\,m\,s$^{-1}$. Therefore,  the Reynolds number, $Re= \dfrac{V \, R \, \rho }{\mu}  \approx 4 \times 10^{-4} \ll 1 $ and the Capillary number, $Ca=\dfrac{V \, R \, \mu }{\rho \, \gamma}  \approx 9  \times 10^{-8} \ll 1 $ which are consistent with our assumptions. 

We fit the experimental data with Eq.~(\ref{eq:balanceG}),
and find that its form is well described with $K_{G}=9.10 \times 10^{-4} \pm 0.1 \times 10^{-4}$\,m\,s$^{-1}$ (see Fig.~\ref{fig:velocityfit1}). If we assume that the drag coefficient for semi-immersed particle $f_d=1$, then from  Eq.~(\ref{eq:forceVella}) we obtain $K_{VM}=\dfrac{C_{VM}}{3 \pi \mu R} = 9.66\times 10^{-4}$\,m\,s$^{-1}$. Considering the significant number of assumptions, the  difference of about 6 $\%$ between the fit and the theory is remarkably small. Further, the decrease of velocity at small particles separation ($l/R < 2.5$ in Fig.~\ref{fig:velocityfit1}) is well described by the Batchelor model of viscous hydrodynamics interactions at low $Re$.

\begin{figure*}
\centering
\includegraphics[width=18cm]{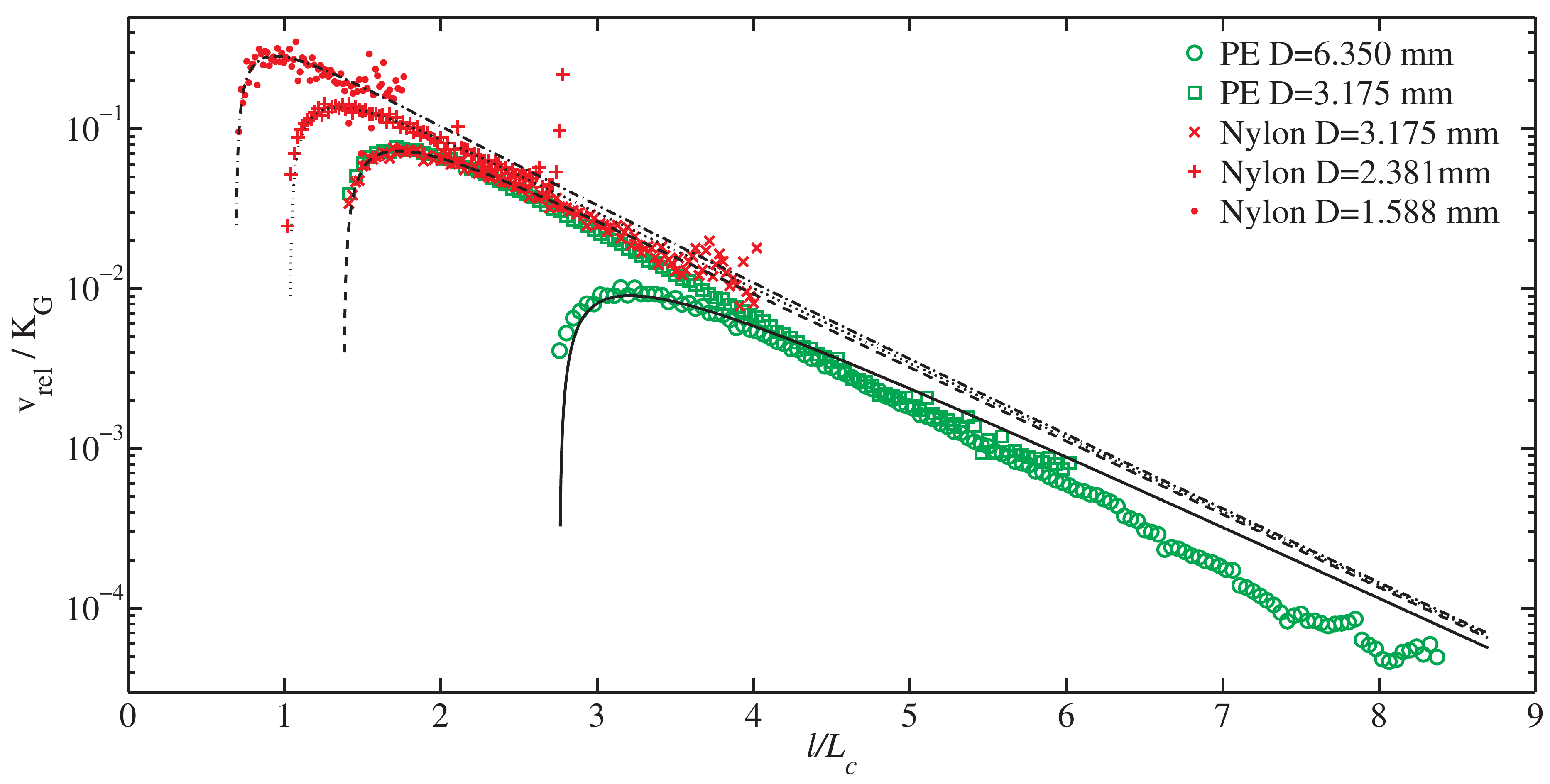}
\caption{Average dimensionless approach velocity  $v_{rel}/K_G$  versus $l/L_c$, for five different kinds of particles listed in Table~\ref{tab:propsphere}. Measurements are compared with the spatial dependency given by Eq.~(\ref{eq:balanceG}).
}
\label{fig:2partalld}
\end{figure*}

\begin{table*}
\centering
\begin{tabular}{p{2.7cm}p{1.5cm}p{3.5cm}p{3.5cm}p{3.6cm}}
\hline
Material & D (mm) &Experimental coefficient $K_{G}$ (m\,s$^{-1}$) & Theoretical coefficient $K_{VM}$ (m\,s$^{-1}$) & Theoretical coefficient $K_{PA}$ (m\,s$^{-1}$)\\ 
\hline \hline
Polyethylene & 6.35  &$2.0\times 10^{-2} \pm 0.3 \times 10^{-2} $ & $3.7\times 10^{-2} \pm 0.7 \times 10^{-2} $ & $6.8\times 10^{-2} \pm 1.2 \times 10^{-2} $\\ 
Polyethylene & 3.175  & $9.1 \times 10^{-4} \pm 0.1 \times 10^{-4}$ & $9.7 \times 10^{-4} \pm 0.7 \times 10^{-4}$ &$11 \times 10^{-4} \pm 0.7 \times 10^{-4}$ \\ 
Nylon & 3.175  & $15\times 10^{-5} \pm 0.1 \times 10^{-5} $ &  $7.9\times 10^{-5} \pm 1.2 \times 10^{-5} $ &$8.1\times 10^{-5} \pm 1.4 \times 10^{-5} $ \\  
Nylon & 2.38125  & $4.2\times 10^{-5} \pm 0.1 \times 10^{-5} $ & $1.8\times 10^{-5} \pm 0.2 \times 10^{-5} $ &$1.9\times 10^{-5} \pm 0.2 \times 10^{-5} $    \\  
Nylon & 1.5875 & $4.5\times 10^{-6} \pm 0.2 \times 10^{-6} $ & $2.4\times 10^{-6} \pm 0.4 \times 10^{-6} $ & $2.4\times 10^{-6} \pm 0.4 \times 10^{-6} $\\ 
\hline 
\end{tabular}
\caption{The coefficients $K_G$ observed in the experiments which sets the amplitude of capillary attraction. The experimental values were obtained using Eq.~(\ref{eq:balanceG}) and errors are estimated from accuracy of the fit. The theoretical values were calculated from the Eqs.~(\ref{eq:forceVassileva}), (\ref{eq:forceVella}) and (\ref{eq:force}), the errors arise from the uncertainties in the parameters values.}
\label{tab:coeff}
  \end{table*}

We measured $K_G$ and $K_{VM}$ with spheres with other sizes and material properties listed Table~\ref{tab:coeff}. The measured velocity is scaled with $K_G$ and compared with Eq.~(\ref{eq:balanceG}) in Fig.~\ref{fig:2partalld}. Good agreement is observed  except for the largest separation for Nylon spheres. 
 The difference may arise due to various reasons. First, Eq.~(\ref{eq:force}) was established assuming that the deformation of the interface and the slope is small. Results for Polyethylene spheres of diameter $D=6.35$\,mm are less in agreement with the model, because we have $z_c/R = 0.154$ and $z_c'=-0.29$, which are not so small compared with $1$.
Second, we assumed $f_d=1$, which may not be accurate. 
Finally, Nylon particles have a higher density, and the resulting capillary force is of an order of magnitude smaller and difficult to measure for large $l/L_c$. 

Overall, our experiments show that capillary attraction between millimetric floating spheres is reasonably well described  by the calculations presented in Sec.~\ref{sec:theory}. For completeness, we present the theoretical results obtained with Eq.~(\ref{eq:forceVassileva}) in Tab.~\ref{tab:coeff}. The agreement is similar to those obtained with Eq.~(\ref{eq:forceVella}) except for the highest $K_G$. While it is not possible to be conclusive as to which model is more accurate, it appears that the Nicolson approximation~\cite{Chan,Vella} is simpler to apply to experiments. Further, our experiments show that both models remain valid for $B_0 \sim 1$ at least when the liquid surface is not strongly distorted  even during contact as in our case where particle centers are well below the liquid surface. 

Our results lead to interesting information on the aggregation process. First, we can estimate the aggregation time using the equation of motion when we neglect hydrodynamics interactions. This approximation is reasonable because the time scale of the approach is dominated by the time when particles are far from each other when viscous hydrodynamic interaction between particles can be neglected (see Fig.~\ref{fig:distancevstime3mm}). If one has a system of isolated spheres, and take the initial separation equal to the mean separation of the particles $l_m$, we estimate the contact time:
\begin{equation}
t_c=\int_{l_m}^{2R} \! \dfrac{-1}{K_G \, K_1(l/L_c)} \, dl.
\end{equation} 
These experiments were performed with $10$ particles ($D=3.175$\,mm), with an initial separation $l_m= 1.85$\,cm. We find that $t_c=13000$\,s, where as 
the estimate gives $t_c(l_m)=24000$\,s which correctly captures its order of magnitude.

Moreover the cohesive force inside an aggregate can be estimated using Eq.~(\ref{eq:forceVella}), with $l=2R$. Because the contact point between the spheres is below the surface, there is indeed no discontinuity of the interface. For spheres with $D=3.175$\,mm, and using the corresponding value of $K_G$, we find the cohesive force at contact: 
$$F_{C}(2R) = 3 \pi \, \mu \, R \, K_G \, K_1(2 R/ L_c) \approx 7.9 \times 10^{-7} {\rm N},  $$
which is significantly smaller than the effective weight of a particle $\approx 4.3 \times 10^{-5}$\,N.
Finally the good agreement between measurements and theoretical results obtained with Nicolson approximation implies that the hydrostatic pressure contribution is negligible which is not obvious when $B_o \sim 1$. Force due to hydrostatic pressure results from the difference of liquid level around the particle caused by the meniscus created by a second particle. Integration of the pressure on a sphere with a spatially varying contact line cannot be accomplished analytically. Using linear superposition of surface deformations and expression of the capillary force between two plates~\cite{Poynting,Vella}, an estimate of the force felt by the first particle due to hydrostatic pressure can be obtained as:
$$F_{P}(2R) \approx \rho \, g \, R \left(  {z_2(3R)}^2- {z_2(R)}^2 \right)  $$
where $z_2(r)$, the deformation of the liquid surface due to the second particle only is computed using Eq.~(\ref{eq:surface}). For $D=3.175$\,mm, we find $F_{P}(2R) \approx 1.3 \times 10^{-7} {\rm N}$. For the largest spheres with  $D=6.35$\,mm, $F_{P}(2R) \approx 4.9 \times 10^{-6} {\rm N}$ when the capillary force becomes $F_{C}(2R) \approx 5.6 \times 10^{-6} {\rm N}$.
Consequently it appears that for particles of similar or larger size, hydrostatic pressure contribution should be taken in account to compute amplitude of capillary attraction.

\section{Experiments with three particles}
\label{subsec:3parts}
\begin{figure}[h]
\centering
\includegraphics[width=7.0cm]{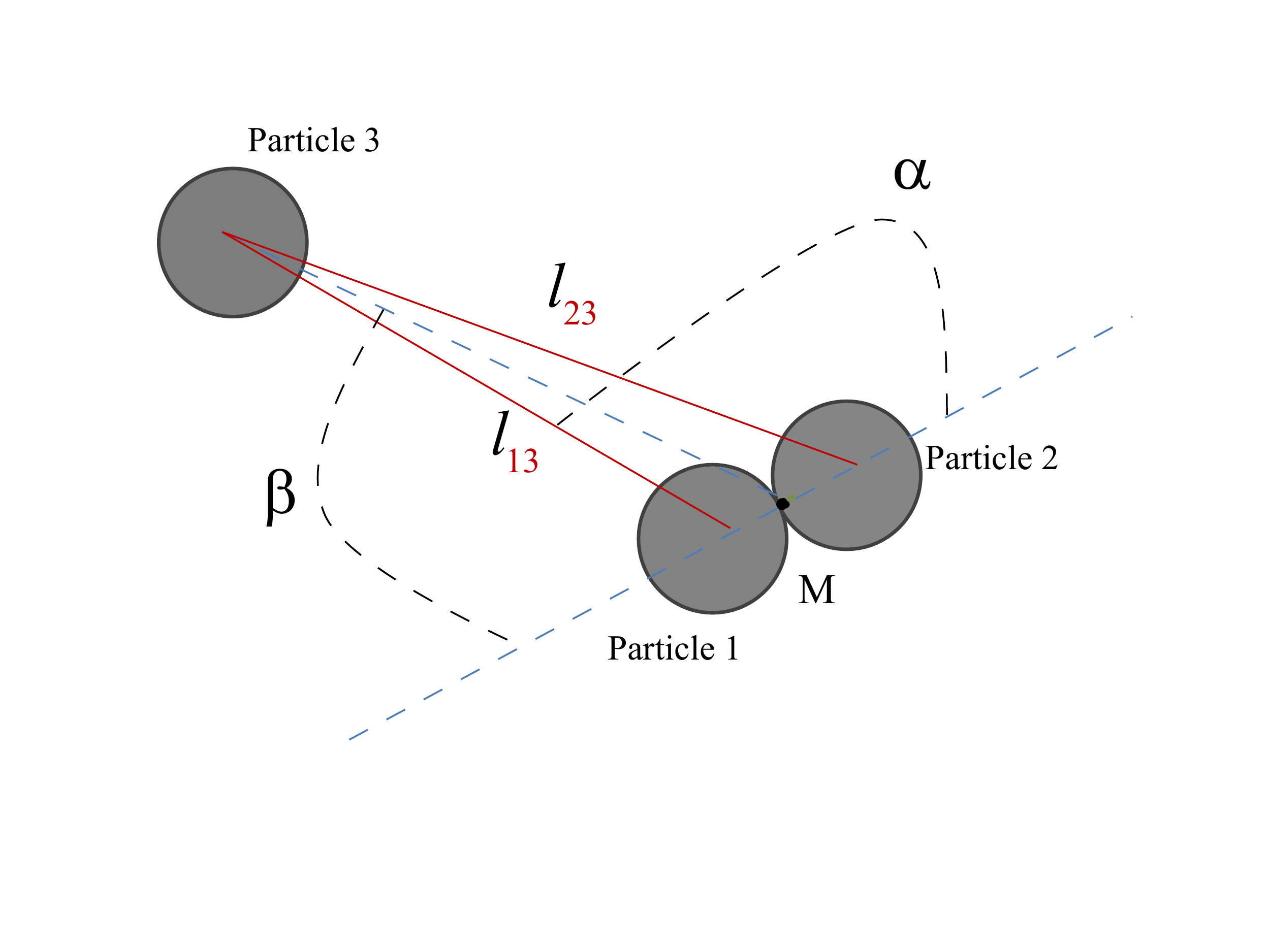}
\caption{The angle $\beta$ and $\alpha$ corresponding to the three particle system viewed from above.
}
\label{fig:3part}
\end{figure}
\begin{figure}
\centering
\includegraphics[width=8.5cm]{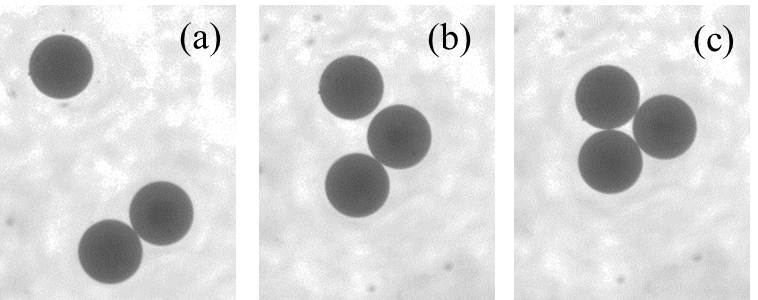}
\caption{A sequence of images showing aggregation of three particles $\beta_i = 77$ degrees and $\beta_f = 90$ degrees. }
\label{fig:3partb}
\end{figure}

We now discuss the formation of aggregates when a third sphere is added at various positions relative to the two sphere cluster. We label particle $1$ and $2$ as the ones that are initially in contact, with the first being closer to the third particle. We label $l_M$ as the distance between the third particle, and $M$, the point of contact between particle $1$ and $2$, and $l_{ij}$ is the distance between particles $i$ and $j$. Inside the cluster, we have $l_{12}=2R=D$. $\beta$ is the angle between particle $1$ and $2$ and particle $3$ and $M$ (see Fig.~\ref{fig:3part}). After the particles are placed in their initial positions, we observe that the cluster of two particles moves as a rigid body and can rotate depending on the initial value of $\beta$. Once contact occurs between the cluster of two particles and the third particle, a rearrangement can occur depending on initial conditions producing an equilateral triangle (Fig.~\ref{fig:3partb}). 

\begin{figure}
\centering
\includegraphics[width=8.5cm]{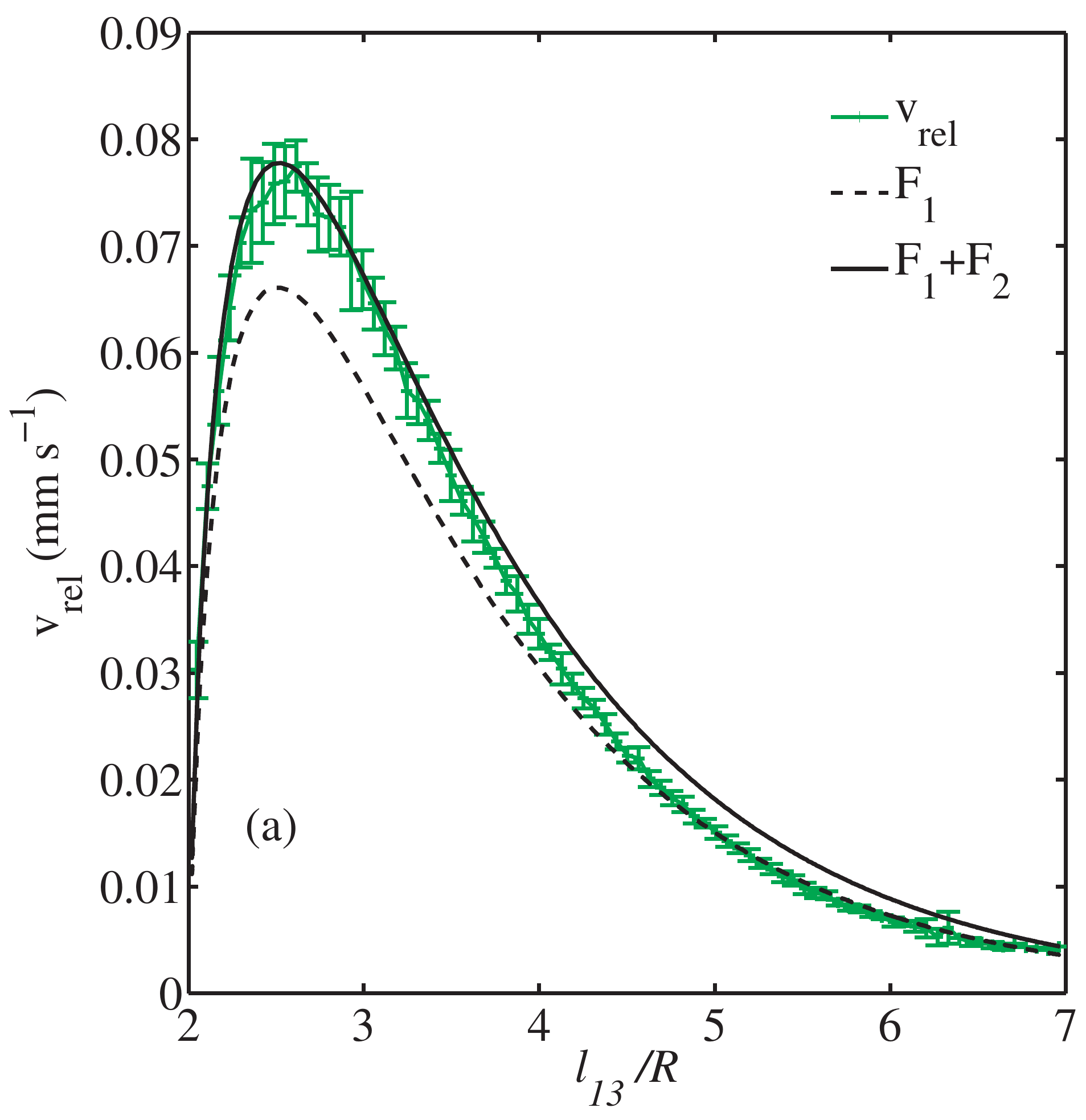}
\includegraphics[width=8.5cm]{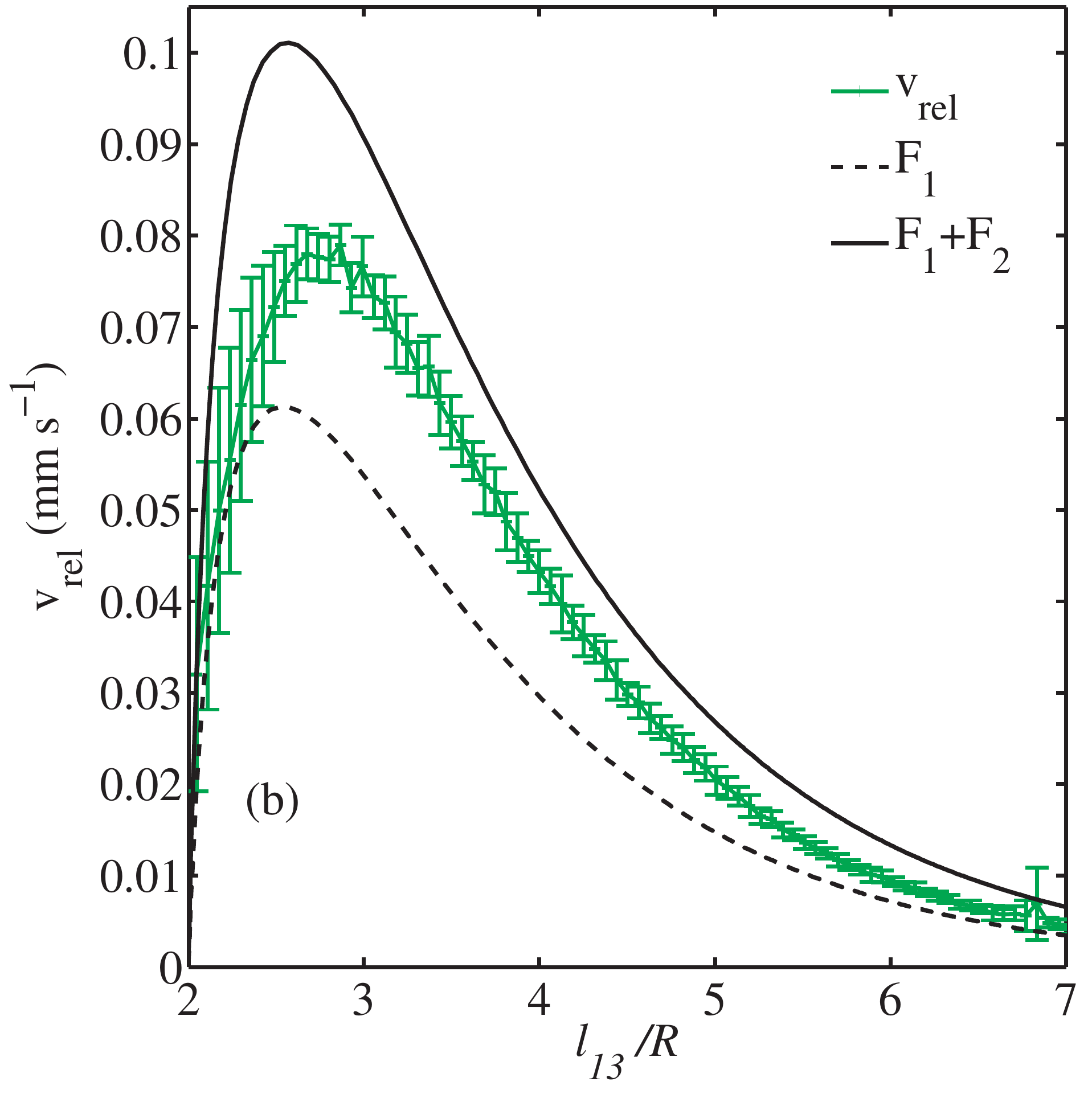}
\caption{Test of the linear superposition assumption for $\beta=0$\,degrees (a) and $\beta=90$\,degrees (b). Average approach velocity of the third particle is plotted as a function of the distance $l$.  
 }
\label{fig:linear}
\end{figure}

\begin{figure}
\centering
\includegraphics[width=8.0cm]{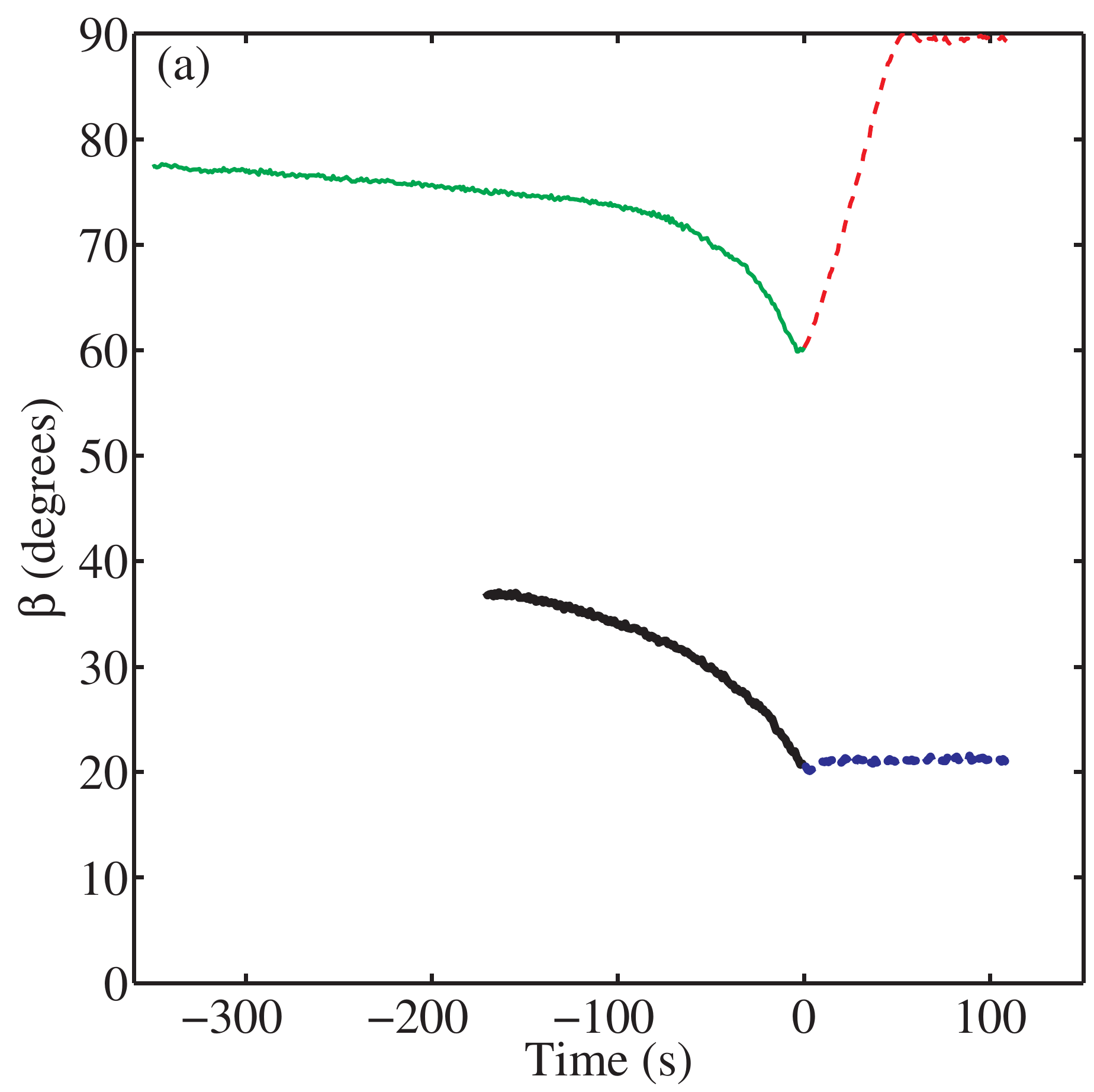}
\includegraphics[width=8.5cm]{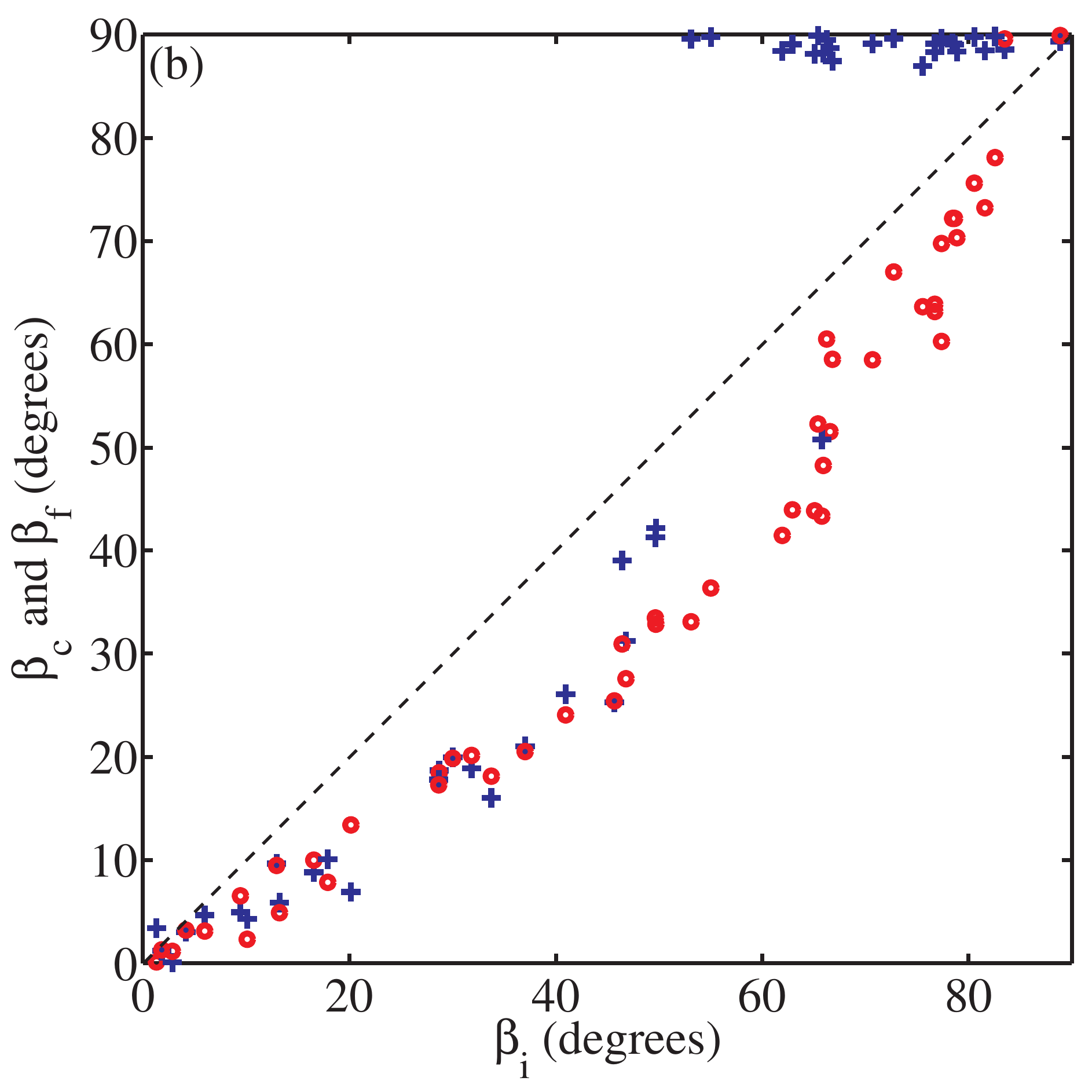}
\caption{(a) Evolution of the angle $\beta$ for polyethylene spheres ($D=3.175$\,mm) with two different initial values, ${77}^\circ$ and  $37^\circ$. The cluster always turns before contact. For large values of $\beta_i$, it also turns after contact, and this evolution is faster than before contact. (b) The angle between the third particle and the center of the cluster of two particles at the instant of contact  $\beta_c$  (red circles) and after any rearrangement as occurred $\beta_f$ (blue crosses) as a function of its initial value $\beta_i$.}
\label{fig:angle}
\end{figure}

\subsection{Capillary force due to a cluster of two particles on a third particle before contact}
We define  $F_{c}$ as the force exerted on the particle $3$ by the two particle cluster (see Fig.~\ref{fig:3part}).  
We have ${\bf F_1}=C\, K_1(l_{13}/L_c)$\,${\bf u_1}$ and ${\bf F_2 }=C\, K_1(l_{23}/L_c)$\,${\bf u_2}$, corresponding to forces which would be exerted by particles $1$ and $2$ respectively on particle $3$, ignoring the effect of the other particle, and ${\bf u_1}$ and ${\bf u_2}$ are associated unitary vectors. 
The approach velocity is defined $\dfrac{d l_M}{d t}$. We consider the projection of  ${\bf F_c}$ along this direction using an unitary vector ${\bf u}$. Then assuming linear superposition, we obtain:
 $${\bf F_c} \cdot{\bf u}= {\bf F_1} \cdot {\bf u}+{\bf F_2}\cdot {\bf u}\,.$$
Drag forces are difficult to estimate in presence of a cluster of two particles. As hydrodynamic interactions are dominated by the contribution of the closest particle, hydrodynamic mobility can be estimated as $G(l_{13}/R)$ in first approximation. Then, 
 \begin{eqnarray}
-dl_M/dt & =  & K_G \, G(l_{13}/R) \, K_1(l_{13}/L_c)\,{\bf u_1}\cdot {\bf u} \nonumber \\
& \quad \quad + & K_G  \, G(l_{13}/R) \, K_1(l_{23}/L_c)\,{\bf u_2}\cdot {\bf u}\,\, .
\label{eq:linear}
\end{eqnarray}
If $l_M/D \gg 1$, then we have $K_1(l_M/L_c) \approx \sqrt{\dfrac{\pi \cdot L_c}{2l}}e^{-l_M/L_c}$. Therefore, $F_1$ and $F_2$ decrease rapidly with separation distance between particles. 
As a result, the closer particle is attracted more strongly which causes the cluster to rotate and decreases $\beta$. 

Next, we test if linear superposition given by Eq.~(\ref{eq:linear}) can be used to describe the formation of the cluster.
The approach velocity $v_{rel}$ is plotted versus the distance between particle $1$ and $3$ $l_{13}$ in Fig.~\ref{fig:linear} for $\beta(t=0)=\beta_i= 0$ degrees (a) and  $\beta(t=0)=\beta_i = 90$ degrees (b). The curves are quite similar to the one obtained previously for attraction between two particles (Fig.~\ref{fig:velocityfit1}). Then the right part of Eq.~(\ref{eq:linear}) and the part exerted by particle 1 only, are added in Fig.~\ref{fig:linear}. We observe that the force exerted by particle $1$ dominates as $l_{13} < l_{23}$ but is not sufficient to describe the total force (Fig.~\ref{fig:linear}). When particles are initially aligned ($\beta_i=0$ degrees (a)), the linear superposition is a good approximation. Decrease of $v_{rel}$ close to contact due to lubrication is also well reproduced using mobility for the closest particle. For a larger initial angle $\beta_i$, result of linear superposition fails to follow experimental curve when the third particle approaches the cluster. Failure of linear superposition may not be surprising because the deformation for two particles is not equal to the sum of the deformation of two individual spheres when the spheres are close to each other. Nevertheless, the linear superposition gives the correct qualitative description, and even gives a reasonable estimate of $F_{c}$ when a particle is at a sufficiently large  distance from the two-sphere cluster.

\subsection{Particle Friction: Shape and dynamics after contact}
We plot the evolution of $\beta$ for two initial angles $\beta_i$ in Fig.~\ref{fig:angle}(a) to illustrate the two different kinds of clusters observed. In both cases, $\beta$ decreases before contact, but while the newly formed 3-particle cluster stops evolving for small initial angles, a rapid rearrangement is observed where $\beta$ increases to 90 degrees for the larger initial angle. This second situation corresponds to formation of an equilateral triangle which is the lowest energy state for a floating three particle cluster.   
In order to investigate the dependence of the two final states on initial conditions, experiments were performed systematically as a function of $\beta_i$ and are represented in Fig.~\ref{fig:angle}(b) by plotting the angle $\beta_{c}$ at contact and $\beta_{f}$ after any rearrangement has occurred. 

As argued previously, the rapid decrease of force with distance tends to decrease $\beta$ before contact as the spheres approach each other. 
For the smallest and largest values of $\beta_i$, cluster do not rotate significantly and $\beta_{c}$ remains very close to $\beta_i$. Then after contact, if $\beta_{i}$ is smaller than $\beta_{i,t}=62\, \pm 2$ degree, the shape of the three particle cluster does not evolve, and thus $\beta_{f}=\beta_{c}$. Otherwise capillary attraction inside the cluster modifies its shape and tends to increase the final angle to $90$ degree. This shape corresponds to a regular triangular cluster where all the spheres are in contact with each other. As the distance between particles are minimal, this state has the lowest possible energy. 

We explain the difference of behavior because of the presence of friction which prevents the spheres from rolling and sliding on each other once they come in contact. Figure~\ref{fig:angle}(b) can be considered as a transition diagram between two phases: compact and aligned clusters due to the competition between friction and capillarity. Below a certain value $\beta_{i,t}$ of $\beta$, final cluster shape is aligned, and the final shape is compact above this value. The few exceptions observed may be due to particle surface irregularities like small bubbles or dust trapped on the spheres.
 
It appears pertinent to express the transition at the contact instant, which gives $\beta_{c,t}=33 \pm 1$ degree and relate this value to the angle $\alpha$ between $l_{12}$ and $l_{13}$ (see Fig.~\ref{fig:3part}). The transition occurs when $\alpha$ is below $\alpha_{c,t}=131 \pm 1$  degrees, when friction can no longer balance the capillary force. Similar behavior is observed using other kinds of particles with different transitions angles $\alpha_{c,t}$ which are reported in Table~\ref{tab:angles}. 
\begin{table}[t]
\centering
\begin{tabular}{cccccc}
\hline
Material & D&  $\beta_{c,t}$  & $\alpha_{c,t}$ & $k$  \\
 & (mm) & (degrees) & (degrees) &  & \\ \hline \hline
Polyethylene &  6.35 &  $41 \pm 4$ & $119 \pm 4$ & $0.0462$   \\  
Polyethylene &  3.175 &  $33 \pm 1$ & $131 \pm 1$ & $0.0761$ \\ 
\hline 
Nylon &  3.175 & $21 \pm 1$& $149 \pm 1$ & $0.0417$  \\  
Nylon &  2.38125 & $29 \pm 3$ & $136 \pm 3$ & $0.0819$  \\ 
Nylon &  1.5875 & $35 \pm 2$ & $128 \pm 2$ & $0.124$ \\ \hline 
\end{tabular}
\caption{Transition angles at contact when triangular compact cluster form, and the effective coefficient of friction for the various kinds of spheres investigated.}
\label{tab:angles}
  \end{table}
Using the transition angles $\alpha_{c,t}$, one can estimate the effective friction coefficient $k$ between particles using the ratio of tangential and normal component of force acting on particle 3 at the point of contact with particle 1. The tangential component is entirely due to force $F_{23}$ due to particle 2, and is given by $F_{23}\cos(\alpha_{c,t}/2)$. The normal component is due to the sum of force $F_{13}$, and component $F_{23}\sin(\alpha_{c,t}/2)$ parallel with $F_{13}$. Therefore,
$$k = \dfrac{F_{23} \cos(\alpha_{c,t} /2) }{F_{13} + F_{23} \sin(\alpha_{c,t} /2)   }. $$
Because $F_{13}= C  \, K_1(D/L_c) $ and $F_{23}= C  \, K_1(l_{23}/L_c) $,  we obtain
\begin{equation}
k \, = \dfrac{K_1(l_{23}/L_c) \cos(\alpha_{c,t} /2) }{K_1(D/L_c) + K_1(l_{23} / L_c) \sin(\alpha_{c,t} /2) }.
\end{equation}
Where, $l_{23}= D\sqrt{2(1-\cos \alpha_{c,t})}$ when particles 1 and 3 are in contact. Computed values of $k$ are reported in Table~\ref{tab:angles}. It may be noted that $k$ decreases with particle size and further depends on the material. While rolling and sliding friction coefficients for these materials on various hard surfaces have been reported and depend on normal force~\cite{ludema}, we were unable to find corresponding data to compare to when particles are immersed in a liquid which may be expected to modify the friction at contact.  

\subsection{Linear chain assembly with capillary and friction forces}
\begin{figure}[t]
\centering
\includegraphics[width=8.5cm]{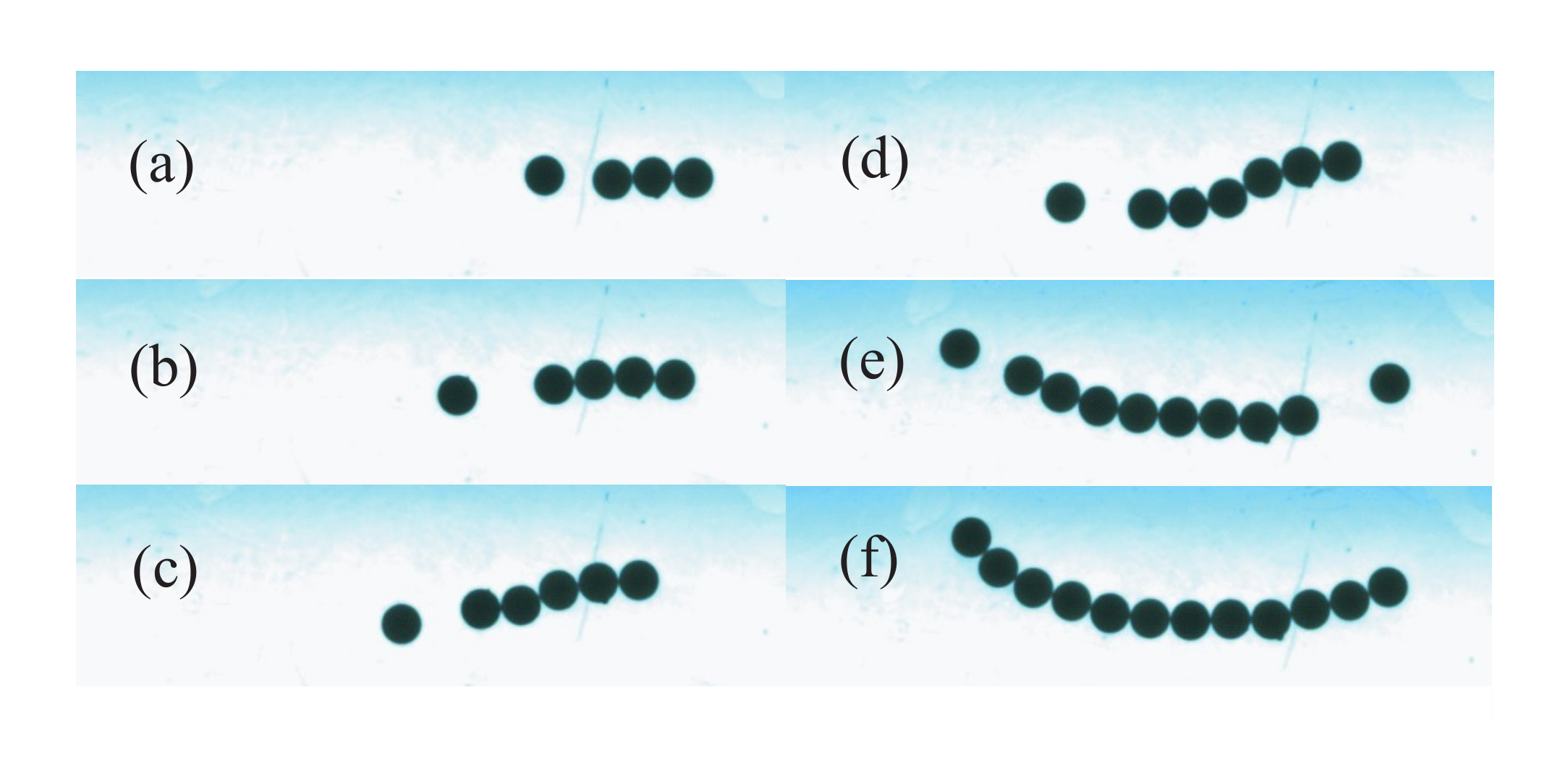}
\caption{(a-f) An example of a chain of twelve spheres obtained by adding successively particles along the cluster axis. The particles are assembled using capillary forces and stabilized by friction. 
\label{fig:chain}
}
\end{figure}

Building on our observation, we demonstrate an example of an aggregate with frictional particles self-organized with capillary forces which is not possible with frictionless particles. The chain is grown initially one particle at a time in Fig.~\ref{fig:chain}(a-d) by dropping a particle near one end of a the developing chain. Then, the chain can be straightened to remove any kinks by introducing particles near opposite ends which tends to straighten the chain (see Fig.~\ref{fig:chain}(e)). The images shown here correspond to polyethylene spheres with $D = 6.35$\,mm. Using this technique, we were able to grow chains as long as 12 particles as shown in Fig.~\ref{fig:chain}(f). Longer chains tend to be unstable and fold to form two rows of particles which is energetically more favorable. 

\section{Conclusions}
\label{sec:Conclusion}
We have experimentally investigated capillary aggregation of small number of floating spheres and shown that attraction between two spheres is reasonably well described by expression of capillary force originally developed for small Bond number $B_o$~\cite{Vella}. Specifically, we have shown that for particles which are hydrophilic and lighter than the fluid, interface deformation and its slope remain small even when Bond number is of order one, allowing one to extend the range of validity of the theory. 
Further, our measurements are sufficiently accurate to also demonstrate the crucial role of hydrodynamic interactions producing rapid decrease of spheres velocity as they come in contact due to viscous effects. 

We have also studied aggregation of a third particle after formation of a cluster of two particles. The theoretical estimate using linear superposition appears to be less accurate in this case, but can be still used to explain the observations qualitatively. The final shape of cluster is determined by the initial position of the third particle relatively to the cluster of two particles. Rotation of the cluster, due to the differential attraction of the two sphere tends to create aligned clusters of three particles. But if the angle $\beta$ between the three particles is small enough, capillary forces between the two particles at the ends overcomes friction to form a compact triangular cluster.

Finally, we note that the observation of large heterogeneous structures in 
larger capillary aggregates~\cite{Berhanu} can be also explained by our analysis. Porous clusters were initially observed to evolve in that study from dilute concentration of particles. As illustrated in our study, particles will rearrange to form a compact triangular cluster only if friction at contact can be overcome by capillary forces. Thus chain like structures which connect various parts of the aggregate can be formed that encompass pores which are locked in place unless compression is applied. Even as particle concentration is increased, such regions lead to defects in the more dense aggregates that form. Our observations demonstrate that friction is clearly necessary to explain shapes of capillary aggregates at the millimeter scale. 

\begin{acknowledgments}
This work was partially supported by National Science Foundation under grant number DMR-0605664. and the donors of the Petroleum Research Fund.
\end{acknowledgments}

\bibliography{dalbe}

\end{document}